\documentstyle[12pt]{article}

\newcommand{\beqn}{\begin{equation}}
\newcommand{\eeqn}{\end{equation}}
\newcommand{\beqnarray}{\begin{eqnarray}}
\newcommand{\eeqnarray}{\end{eqnarray}}

\newcommand{\rd}{\partial}

\newcommand{\bfZ}{{\bf Z}}
\newcommand{\calF}{{\cal F}}
\newcommand{\calM}{{\cal M}}

\newcommand{\calP}{{\cal P}}
\newcommand{\calT}{{\cal T}}
\newcommand{\Stilde}{{\tilde S}}
\newcommand{\Ttilde}{\tilde{T}}
\newcommand{\Xtilde}{{\tilde X}}
\newcommand{\Omegatilde}{\tilde{\Omega}}
\newcommand{\Omegahat}{\hat{\Omega}}
\newcommand{\abar}{\bar{a}}
\begin{document}

\title{Integrable Hierarchies and Contact Terms in 
u-plane Integrals of Topologically Twisted Supersymmetric 
Gauge Theories}
\author{Kanehisa Takasaki\\
{\normalsize Department of Fundamental Sciences, Kyoto University}\\
{\normalsize Yoshida, Sakyo-ku, Kyoto 606, Japan}\\
{\normalsize E-mail: \tt takasaki@yukawa.kyoto-u.ac.jp}}
\date{}
\maketitle

\begin{abstract} 
\noindent
The $u$-plane integrals of topologically twisted $N = 2$ supersymmetric 
gauge theories generally contain contact terms of nonlocal topological 
observables.  This paper proposes an interpretation of these contact 
terms from the point of view of integrable hierarchies and their 
Whitham deformations.  This is inspired by Mari\~no and Moore's remark 
that the blowup formula of the $u$-plane integral contains a piece
that can be interpreted as a single-time tau function of an integrable 
hierarchy.  This single-time tau function can be extended to a 
multi-time version without spoiling the modular invariance of the 
blowup formula.  The multi-time tau function is comprised of a Gaussian 
factor $e^{Q(t_1,t_2,\ldots)}$  and a theta function.  The time 
variables $t_n$ play the role of physical coupling constants of 
2-observables $I_n(B)$ carried by the exceptional divisor $B$.  The 
coefficients $q_{mn}$ of the Gaussian part are identified to be the 
contact terms of these 2-observables.  This identification is further
examined in the language of Whitham equations.  All relevant quantities 
are written in the form of derivatives of the prepotential. 
\end{abstract}
\bigskip

\begin{flushleft}
KUCP-0115\\
hep-th/9803217
\end{flushleft}
\newpage
%%%%%%%%%%%%%%%%%%%%%%%%%%%%%%%%%%%%%%%%%%%%%%%%%%%%%%%%%%%%%%%%%%%%

\section{Introduction}

The ``$u$-plane integral'' of Moore and Witten \cite{bib:Moore-Witten} 
gives an exact answer to the contribution of the Coulomb branch 
in correlation functions of topologically twisted four-dimensional 
$N = 2$ supersymmetric gauge theories.  
Moore and Witten considered the case of $SU(2)$ and $SO(3)$ only.  
Mari\~no and Moore \cite{bib:Marino-Moore} extended the $u$-plain 
integral to more general, higher rank gauge groups.  Losev et al. 
studied the $u$-plane \cite{bib:Losev-etal} integral from 
a somewhat different point of view (``Gromov-Witten paradigm'').   

The correlation functions of those topologically twisted 
$N = 2$ supersymmetric gauge theories are the Donaldson-Witten 
invariants of the four-manifold $X$ \cite{bib:Donaldson-Witten}.  
Their generating function is given by a path integral of the form 
\beqn
    Z_{\rm DW}(xS + yP) = \Bigl< \exp( xI(S) + yO(P) ) \Bigr> 
\eeqn
with book-keeping variables $x$ and $y$ coupled to the 2-cycle 
$S \in H_2(X,\bfZ)$  and the 0-cycle $O \in H_0(X,\bfZ)$. 
$I(S)$ and $O(P)$ are associated observables (``2-observable'' 
and ``0-observable'').   The Coulomb branch has a nonvanishing 
contribution only if $b_2^+ \le 1$, and this contribution is 
given by an integral of the following form (``u-plane integral'') 
over the Coulomb moduli space: 
\beqn
   Z_u = \int_{\calM_{\rm Coulomb}} da d\abar A(u)^\chi B(u)^\sigma 
         \exp(U + S^2 T(u)) \Psi. 
\eeqn
$\chi$ and $\sigma$ are respectively the Euler characteristic and 
the signature of $X$; $U$ is a contribution of the 0-observable; 
$T$ is the `` contact term'' induced by the 2-observable; $\Psi$ 
is the photon partition function, which is a lattice sum of the 
Siegel-Narain type over the tensor product of $H_2(X,\bfZ)$ and 
the weight lattice of the gauge group $G$.  Since the geometry 
of the Coulomb moduli space is determined by the low energy 
effective action of Seiberg and Witten (``Seiberg-Witten 
solution'') \cite{bib:Seiberg-Witten}, this $u$-plane integral 
gives a complete answer to the physics of the Coulomb branch 
at least in the topologically twisted theories.

The low energy effective action of the Coulomb branch is 
generally related to an integrable Hamiltonian system 
\cite{bib:SW-IHS}.  Our subsequent consideration is mostly 
focussed on the $SU(N)$ theory without matter hypermultiplet 
($N_f = 0$) \cite{bib:SW-SU(N)}.  In this case, 
the integrable Hamiltonian system is the $N$-periodic 
Toda chain.  Its spectral curve can be written 
\beqn
    z^2 - \Lambda^{-N} P(x) z + 1 = 0, 
    \label{eq:Toda-curve}
\eeqn
where $P(x)$ is a polynomial of the form 
\beqn
    P(x) = x^N - \sum_{j=2}^N u_j x^{N-j}. 
\eeqn
This hyperelliptic curve of genus $g = N - 1$ now plays the 
role of the Seiberg-Witten elliptic curves in the $SU(2)$ 
theories \cite{bib:Seiberg-Witten}. The coefficients $u_j$, 
which give a Poisson-commuting set of Hamiltonians of the 
Toda chain, are thereby identified with the Coulomb moduli. 
$\Lambda$ is the energy scale of the low energy theory.  
The low energy effective action is written in terms of a 
prepotential $\calF = \calF(a)$, $a = (a_1,\cdots,a_{N-1})$.  
This prepotential is determined (though as an implicit function) 
by the functional relation 
\beqn
    a^D_j = \frac{\rd \calF}{\rd a_j}
\eeqn
of the period integrals 
\beqn
    a_j = \oint_{\alpha_j} dS_{\rm SW}, \quad 
    a^D_j = \oint_{\beta_j} dS_{\rm SW}
\eeqn
of the meromorphic differential 
\beqn
    dS_{\rm SW} = x d \log z 
\eeqn
along a symplectic basis $\alpha_j,\beta_j$ 
($j=1,\cdots,N-1$) of cycles on the Riemann surface 
of the spectral curve.  The differential $dS_{\rm SW}$ 
obeys the fundamental relation 
\beqn
    \rd_{a_j} dS_{\rm SW}|_{z={\rm const}}
        = d\omega_j. 
\eeqn
The left hand side means differentiating $dS_{\rm SW}$ 
while keeping $z$ constant.  $d\omega_j$ ($j= 1,\cdots,N-1$) 
are a basis of holomorphic differentials normalized as
\beqn
    \oint_{\alpha_k} d\omega_j = \delta_{jk}. 
\eeqn
This equation connecting $dS_{\rm SW}$ and $d\omega_j$ 
shows a link with the notion of ``Whitham equations'' 
for adiabatic deformations of algebro-geometric solutions 
of integrable systems.

It is accordingly natural to expect a similar relation of 
the $u$-plane integrals to some integrable systems and 
associated Whitham equations.  Mari\~no and Moore 
\cite{bib:Marino-Moore} and Losev et al. \cite{bib:Losev-etal}, 
independently, presented several interesting remarks towards 
that direction. The purpose of this paper is to examine their 
remarks in more detail.

\section{Blowup formula and contact terms}

One of remarks of Mari\~no and Moore \cite{bib:Marino-Moore} 
is that the ``blowup formula'' of the $u$-plane integrals 
contains a factor that can be interpreted as a special tau 
function of the Toda chain (or, more precisely, the Toda 
lattice hierarchy \cite{bib:TL-hierarchy}).  

The blowup formula is concerned with the blowup $\Xtilde$ 
of the four-manifold $X$ at a point $P$. Let $B$ denote 
the homology class of the exceptional divisor (i.e., 
the inverse image of the blowup point), and take the homology 
class $\Stilde = S + tB$ with a parameter $t$ on $\Xtilde$. 
Furthermore, let the metric of $\Xtilde$ be such that the 
Poincar\'e dual of $B$ is anti-self-dual, i.e., $B_+ = 0$. 
In the case of $G = SU(N)$ and $N_f = 0$, the blowup formula 
shows that the effect of blowup is just to replace the 
0-observable factor $e^U$ as
\beqn
    e^U \to 
    e^U \frac{\alpha}{\beta} 
    \det\left(\frac{\rd u_k}{\rd a_j}\right)^{1/2} \Delta^{-1/8} 
    e^{-t^2T} \Theta_{\gamma,\delta}\Big(\frac{1}{2\pi}tV \mid \calT\Bigr). 
    \label{eq:exp-U} 
\eeqn
Here $\alpha$ and $\beta$ are some numerical constants, $\Delta$ 
the discriminant of the above Toda spectral curve, and the last 
factor is the ordinary theta function ($g = N-1$) 
\beqn
    \Theta_{\gamma,\delta}(w \mid \calT) 
    = \sum_{\ell \in \bfZ^g} 
      \exp\left[ \pi i <\ell + \gamma, \calT(\ell + \gamma)> 
        + 2 \pi i <\ell + \gamma, w + \delta> \right]
\eeqn
with the half-characteristic $(\gamma,\delta)$ determined by 
the setup of the $u$-plane integral.  The period matrix 
$\calT = (\calT_{jk})_{j,k=1,\cdots,g}$ is defined by 
\beqn
    \oint_{\beta_k} d\omega_j = \calT_{jk}. 
\eeqn
The vector $V = (V_j)_{j=1,\cdots,g}$ is a gradient vector, 
with respect to $a = (a_1,\cdots,a_g)$, of a function 
$\calP = \calP(a)$: 
\beqn
    V_j = \frac{\rd \calP}{\rd a_j}. 
\eeqn
This function $\calP$ appears in the definition of the 
2-observable $I(S)$ as 
\beqn
    I(S) = {\rm const.} \int_S G^2 \calP, 
\eeqn
where $G$ is an operator that generates a standard solution 
of the descent equations for observables 
\cite{bib:Moore-Witten,bib:Marino-Moore}.  The contact term $T$ 
also depends on this potential function $\calP$, therefore should 
be written $T_\calP$ more precisely.

It is the last two factors in (\ref{eq:exp-U}) that Mari\~no 
and Moore identified to be a tau function of the Toda lattice 
hierarchy. This interpretation is very suggestive in the 
following sense. 
\begin{itemize}
\item The parameter $t$ is interpreted as a time variable in 
the hierarchy.  This is in sharp contrast with the status of 
the Toda chain in the aforementioned description of the 
Coulomb moduli space.  The role of the Toda chain is simply 
to supply a $g$-dimensional family of curves along with 
a special-geometric structure; the dynamics of the Toda chain, 
as an integrable Hamiltonian system, plays no role.  In the 
$u$-plane integral, meanwhile, the time variable becomes 
a physical coupling constant of the observable $I(B)$. 

\item The contact term $T$ and the directional vector $V$ 
are determined by the potential $\calP$.  In this sense, 
$\calP$ is a Hamiltonian of the Toda lattice hierarchy. 
In fact, any polynomial (or holomorphic function) of 
the Coulomb moduli can be used as $\calP$.  One can 
consider a set of commuting flows with a set of 
Hamiltonians $\calP_1,\calP_2,\cdots$ and associated time 
variables $t_1,t_2,\cdots$.  These commuting flows generally  
form a subhierarchy of the standard full Toda lattice 
hierarchy with two infinite series of time variables 
$t_n^\pm$ ($n = 1,2,\cdots$) \cite{bib:TL-hierarchy}. 
\footnote{Actually, the Toda lattice hierarchy has another 
variable $t_0$ --- the lattice coordinates.  In order to 
avoid unnecessary complication, we shall not consider it, 
or just put $t_0 = 0$.} 
Of course, the notion of tau function is also meaningful 
for such a subhierarchy.   This strongly suggests that 
the blowup formula, too, can be extended in that way. 
\end{itemize}

\section{Insertion of more than one 2-observables} 

Let us specify the implication of the second point above. 
The potentials $\calP_1,\calP_2,\cdots$, determine the 
2-observables 
\beqn
    I_n(B) = {\rm const.} \int_B G^2 \calP_n. 
\eeqn
These 2-observables can be used to deform the correlation 
functions as
\beqn
    \Bigl< \exp\Bigl( I(S) + O(P) \Bigr)\Bigr> \to 
    \Bigl< \exp\Bigl( I(S) + \sum_n t_n I_n(B) + O(P) \Bigr)\Bigr>. 
    \label{eq:deformed-correlator}
\eeqn
(The book-keeping parameters are set to $x = y = 1$.) 
This will modify the last two factors of 
(\ref{eq:exp-U}) as
\beqnarray
    && \exp(-t^2 T)\Theta_{\gamma,\delta}(\frac{1}{2\pi}tV \mid \calT) 
    \nonumber \\
    && \to \exp\Bigl( - \sum_{m,n} C(\calP_m,\calP_n) t_m t_n \Bigr) 
           \Theta_{\gamma,\delta}(\sum_n t_n V_n \mid \calT), 
    \label{eq:deformed-factors}
\eeqnarray
where $C(\calP_m,\calP_n)$ are the two-body contact terms induced 
by the new 2-observables.  Note that since these 2-observables 
are carried by the same 2-cycle $B$, higher contact terms do not 
appear.  The negative sign in front of the contact terms originates 
in the self-intersection number $B^2 = -1$ of the exceptional 
divisor.

Some of these contact terms have been calculated explicitly.  
The simplest case of $C(u_2,u_2)$ for $G = SU(2)$ is due to 
Moore and Witten \cite{bib:Moore-Witten}; the result is 
written in terms of elliptic theta functions and Eisenstein series.  
This result is generalized to $G = SU(N)$ by Mari\~no and Moore 
\cite{bib:Marino-Moore}. 
Their method is based on the so called RG (renormalization 
group) equation \cite{bib:SW-RGE} 
\beqn
    \frac{\rd \calF}{\rd \log\Lambda} = {\rm const.\ } u_2 
\eeqn
and the modular transformations 
\beqnarray
    \frac{\rd^2 \calF}{(\rd \log\Lambda)^2} 
    &\to& 
    \frac{\rd^2 \calF}{(\rd \log\Lambda)^2} 
    - \frac{\rd^2 \calF}{\rd \log\Lambda \rd a_j} 
      [(C\calT + D)^{-1} C]_{jk}
      \frac{\rd^2 \calF}{\rd \log\Lambda \rd a_k}, 
    \nonumber \\
    \frac{\rd^2 \calF}{\rd \log\Lambda \rd a_j} 
    &\to&
    [(C\calT + D)^{-1}]_{jk} \frac{\rd^2 \calF}{\rd \log\Lambda \rd a_k}
\eeqnarray
under the symplectic transformations of cycles
\beqn
    \begin{array}{ll}
    \beta_j \to A_{jk} \beta_k + B_{jk} \alpha_k, \\
    \alpha_j \to C_{jk} \beta_k + D_{jk} \alpha_k, 
    \end{array}
    \quad 
    \left( \begin{array}{ll}
      A & B \\
      C & D
    \end{array} \right)  \in Sp(2g, \bfZ).  
    \label{eq:Sp-on-cycles}
\eeqn
The second quantity above are the components of the directional 
vector $V$: 
\beqn
    V_j = {\rm const.\ } \frac{\rd^2 \calF}{\rd \log\Lambda \rd a_j}.
\eeqn
Mari\~no and Moore thus obtained the following formula: 
\beqn
    C(u_2,u_2) = {\rm const.\ } \frac{\rd^2 \calF}{(\rd \log\Lambda)^2}. 
    \label{eq:Cu2u2-F}
\eeqn
By the results from the RG equation, this formula can be rewritten 
\beqn
    C(u_2,u_2) = \frac{{\rm const.}}{2N - N_f} 
        \left( 2 u_2 - a_j \frac{\rd u_2}{\rd a_j} \right). 
\eeqn
(Here the result is presented in a generalized form with $N_f$ 
massless hypermultiplets).  Losev et al. \cite{bib:Losev-etal} 
derived a more general result: 
\beqn
    C(u_2,u_k) = \frac{{\rm const.}}{2N - N_f} 
        \left( k u_k -  a_j \frac{\rd u_k}{\rd a_j} \right). 
\eeqn
(We have omitted an explicit form of various constants above, 
which are irrelevant in our subsequent analysis.)

\section{Multi-time tau function and contact terms} 

Let us now proceed to our interpretation of the contact 
terms $C(\calP_m,\calP_n)$.  As we have noted above, 
the product of the last two factor in (\ref{eq:exp-U}) 
persists to be a tau function of the Toda lattice hierarchy 
for any choice of the potential $\calP$ of the 2-observable 
$I(B)$.  Meanwhile, the notion of the tau function can be 
readily generalized to the multi-time setting.  Therefore, 
a natural conjecture is that right hand side of 
(\ref{eq:deformed-factors}) will be a multi-time tau 
function of the Toda lattice hierarchy: 
\beqn
    \tau_{\gamma,\delta}(t_1,t_2,\cdots) = 
    \exp\Bigl( - \sum_{m,n} C(\calP_m,\calP_n) t_m t_n \Bigr) 
    \Theta_{\gamma,\delta}(\sum_n t_n V_n \mid \calT). 
    \label{eq:conjecture-tau}
\eeqn

More precisely, this tau function should be an 
algebro-geometric tau function 
\cite{bib:Nakatsu-Takasaki} determined by a 
suitable set of data (the Krichever data) on the 
Toda spectral curve (\ref{eq:Toda-curve}) 
\cite{bib:Krichever-Toda}. 
(We here consider the case of $N_f = 0$ only. 
Other cases can be treated in a similar way.)   
Such a tau function, just like the algebro-geometric 
tau functions of other integrable hierarchies 
\cite{bib:tau-theta}, takes the form 
\beqn
    \tau(t_1,t_2,\cdots) = e^{Q(t_1,t_2,\cdots)} 
        \Theta(\sum_n t_n V_n + w_0 \mid \calT), 
\eeqn
where $e^{Q(t_1,t_2,\cdots)}$ is a Gaussian factor 
(including a linear part), 
\beqn 
    Q(t_1,t_2,\cdots) = \frac{1}{2} \sum_{m,n} q_{mn} t_m t_n 
        + \sum_n r_n t_n + r_0, 
\eeqn
$\Theta(w \mid \calT)$ is the ordinary theta function 
without characteristic, and $w_0$ is a constant vector. 
As we show later, $q_{mn}$ and $V_n$ are determined by 
the the algebro-geometric data.  $r_n$, $r_0$ and $w_0$ 
are arbitrary constants.  Our conjectural tau function 
(\ref{eq:conjecture-tau}) can be rewritten into the 
above form using the relation 
\beqn
    \Theta_{\gamma,\delta}(w \mid \calT) = 
      e^{2 \pi i<\gamma,w + \delta>} 
      \Theta(w + \calT \gamma + \delta \mid \calT) 
\eeqn
between the two theta functions.  In particular, the 
arbitrary constants $r_n$, $r_0$ and $w_0$ are 
determined by the half-characteristic and the period 
matrix, and $q_{mn}$ turn out to be essentially 
the contact terms: 
\beqn
    C(\calP_m, \calP_n) = - \frac{1}{2} q_{mn}. 
    \label{eq:conjecture-CandQ}
\eeqn

The above conjecture is supported by a modular transformation 
property of the tau function $\tau_{\gamma,\delta}(t_1,t_2,\cdots)$ 
under symplectic transformations of cycles (\ref{eq:Sp-on-cycles}). 
Namely, as we shall show below, the tau function turns out to 
possess the same $t$-INDEPENDENT modular property as the original 
$t$-dependent factors in (\ref{eq:exp-U}); this implies that the 
corrected 0-observable factor (\ref{eq:exp-U}) persists to be 
modular invariant after modifying the last two factors as in 
(\ref{eq:deformed-factors}).  This is strong evidence 
(though not a proof) of the correctness of the conjecture.

Let us examine modular transformations of the tau function 
$\tau_{\gamma,\delta}(t_1,t_2,\cdots)$.  Fortunately, this kind 
of problems have been already studied in the context of 
free fermion systems on Riemann surfaces \cite{bib:tau-modular}.  
(This also suggests a possible link of four-dimensional 
supersymmetric gauge theories with two-dimensional conformal 
field theories.)  Remarkably, the modular transformation 
property of the tau function is quite universal, i.e., does 
not depend on the detail of the Riemann surface or the 
algebro-geometric data (even nor the integrable system 
itself!).  In the following, therefore, we consider the 
tau function $\tau_{\gamma,\delta}(t_1,t_2,\cdots)$ for 
an arbitrary compact Riemann surface $\Sigma$ of genus $g$.

The fundamental algebro-geometric data in the case of the 
Toda lattice hierarchy \cite{bib:Nakatsu-Takasaki} 
are comprised of two marked points $P_\infty^\pm$, 
local complex coordinates $z_\pm$ at each of 
those points normalized as $z_\pm(P_\infty^\pm) = 0$, 
and the symplectic basis of cycles $\alpha_j,\beta_j$ 
($j = 1,\cdots,g$).  Furthermore, a set of polynomials 
$f_n^\pm(z_\pm^{-1})$ ($n = 1,2,\cdots$) have to be given 
in order to select the ``directions'' of the $t_n$'s in 
the total space of the standard time variables $t_n^{\pm}$ 
($n = 1,2,\cdots)$ of the Toda lattice hierarchy 
\cite{bib:TL-hierarchy}.  
Each pair $f_n^+(z_\pm^{-1})$ and $f_n^-(z_\pm^{-1})$ 
determines the direction of the $n$-th time (or, equivalently, 
a Hamiltonian $\calP_n$ in the sense already mentioned).  

Given these data, the following meromorphic differentials 
$d\Omega_n$ ($n = 1,2,\cdots$) are uniquely determined: 
\begin{itemize}
\item $d\Omega_n$ is holomorphic everywhere except at 
$P_\infty^\pm$. In a neighborhood of $P_\infty^\pm$, 
respectively, 
\beqn
    d\Omega_n = df_n^\pm(z_\pm^{-1}) + {\rm holomorphic}. 
\eeqn
\item The $\alpha$-periods all vanish, 
\beqn
    \oint_{\alpha_j} d\Omega_n = 0 \quad (j = 1,\cdots,g). 
\eeqn
\end{itemize}

These meromorphic differentials determine $q_{mn}$ and $V_n$ 
as follows. The components of the vector $V_n$ are given by 
\beqn
    V_{jn} = \frac{1}{2\pi i} \oint_{\beta_j} d\Omega_n. 
\eeqn
By Riemann's bilinear relation, this can be rewritten 
\beqn
    V_{jn} = 
      - \frac{1}{2\pi i} \oint_{P_\infty^+} f_n^+(z_+^{-1}) d\omega_j  
      - \frac{1}{2\pi i} \oint_{P_\infty^-} f_n^-(z_-^{-1}) d\omega_j, 
    \label{eq:Vjn-integral}
\eeqn
where the integrals on the right hand side are a contour integral 
turning once around $P_n^\pm$ respectively. $q_{mn}$ is given by 
replacing $d\omega_j$ by $d\Omega_n$: 
\beqn
    q_{mn} = 
      - \frac{1}{2\pi i} \oint_{P_\infty^+} f_n^+(z_+^{-1}) d\Omega_m
      - \frac{1}{2\pi i} \oint_{P_\infty^-} f_n^-(z_-^{-1}) d\Omega_m. 
    \label{eq:qmn-integral}
\eeqn
These somewhat complicated expressions are a linear combination 
of more familiar expressions of the $q$'s and $V$'s for the 
standard time variables $t_n^\pm$ \cite{bib:Nakatsu-Takasaki}.

The modular transformation properties of these quantities 
can be derived by straightforward calculations.  Under the 
symplectic transformation of cycles (\ref{eq:Sp-on-cycles}), 
the holomorphic and meromorphic differentials transform as 
\beqnarray
    d\omega_j &\to& [(C\calT + D)^{-1}]_{kj} d\omega_k, 
    \nonumber \\
    d\Omega_n &\to& 
    d\Omega_n - [(C\calT + D)^{-1} C]_{k\ell} 
    \left(\oint_{\beta_\ell} d\Omega_n \right) d\omega_k. 
\eeqnarray
Accordingly, $V_n$ and $q_{mn}$ transform as
\beqnarray
    V_n &\to& {}^t (C\calT + D)^{-1} V_n, 
    \nonumber \\
    q_{mn} &\to& 
    q_{mn} - 2 \pi i {}^tV_m (C\calT + D)^{-1} C V_n. 
\eeqnarray

The final piece of the ring is the following modular transformation 
formula of theta functions \cite{bib:Mumford}: 
\beqnarray
    && \Theta_{\gamma,\delta}\Bigl( {}^t(C\calT + D)^{-1} w \mid 
         (A\calT + B)(C\calT + D)^{-1} \Bigr) 
    \nonumber \\
    && = \epsilon \det(C\calT + D)^{1/2} 
      \exp( \pi <w, {}^t(C\calT + D)^{-1}w> )
      \Theta_{\gamma',\delta'}(w \mid \calT).  
\eeqnarray
Here $\epsilon$ is an $8^{\rm th}$ root of unity, 
$\epsilon^8 = 1$, and $(\gamma',\delta')$ is a new 
half-characteristic, both determined by the symplectic matrix.   
Combining this formula with the above modular transformations 
of $q_{mn}$ and $V_n$ lead to the conclusion that the tau function 
$\tau_{\gamma,\delta}(t_1,t_2,\cdots)$ has a $t$-independent 
modular transformation property: 
\beqn
    \tau_{\gamma,\delta}(t_1,t_2,\cdots) 
      \to 
    \epsilon \det(C\calT + D)^{1/2} 
      \tau_{\gamma',\delta'}(t_1,t_2,\cdots). 
\eeqn
Understanding that the half-characteristic is also 
transformed because of its physical origin, one can thus 
confirm that the tau function in our conjecture possess 
a desirable modular transformation 
property.

\section{Perspectives from Whitham equations and prepotential} 

Let us reconsider the meaning of $q_{mn}$, $V_{jn}$ and 
$\calP_n$ in the language of Whitham deformations of the 
integrable hierarchy.  The most suggestive in this respect 
is the formula (\ref{eq:Cu2u2-F}) of $C(u_2,u_2)$ 
as a second derivative of the prepotential. 
The relation between RG equations and Whitham equations 
\cite{bib:SW-RGE} tells us that, very roughly speaking, 
the energy scale $\Lambda$ can be identified with the first 
Whitham time variable $T_1$.  In view of this fact, a natural 
conjecture is that the contact terms of higher observables 
can be written 
\beqn
    C(\calP_m, \calP_n) = 
      {\rm const.\ } \frac{\rd^2 \calF}{\rd T_m \rd T_n}. 
\eeqn
This conjecture indeed turns out to fit into the general 
framework of Whitham equations proposed in our previous work 
\cite{bib:Nakatsu-Takasaki}, as we show below.

The Whitham equations in the present situation take the form 
\beqn
    \rd_{T_n} dS|_{z={\rm const.}} = d\Omega_n, \quad 
    \rd_{a_j} dS|_{z={\rm const.}} = d\omega_j, 
    \label{eq:Whitham}
\eeqn
where $dS$ is given by 
\beqn
    dS = \sum_n T_n d\Omega_n + \sum_{j=1}^{N-1} a_j d\omega_j, 
    \label{eq:dS-condition}
\eeqn
and ``$|_{z={\rm const.}}$'', also here, means differentiating 
while keeping $z$ constant.  These equations give deformations 
of the Coulomb moduli $u_j = u_j(a,T)$ as the Whitham time 
variables $T = (T_1,T_2,\cdots)$  vary from a point of departure 
(e.g., a point $T = T_{\rm SW}$ where $dS$ coincides with the 
meromorphic differential $dS_{\rm SW}$). One can redefine 
the prepotential $\calF = \calF(a,T)$, now as a function of 
$a$ and $T$, by the equations
\beqn
    \frac{\rd \calF}{\rd T_n} = 
      - \oint_{P_\infty^+} f_n^+(z_+^{-1}) dS 
      - \oint_{P_\infty^-} f_n^-(z_-^{-1}) dS 
    \label{eq:DefEqF-Tn}
\eeqn
and 
\beqn
    \frac{\rd \calF}{\rd a_j} = \oint_{\beta_j} dS. 
    \label{eq:DefEqF-aj}
\eeqn
Recall that $f_n^\pm(z_\pm^{-1})$ are the polynomials giving 
the singular part of $d\Omega_n$ at $P_\infty^\pm$. 
The compatibility (integrability) of the above equations 
for $\calF$ is again a consequence of Riemann's bilinear relation.

Now differentiate the right hand side of the defining equation 
of $\rd \calF/\rd T_n$ against $a_j$ and $T_m$.  By the Whitham 
equations, the outcome is nothing but the right hand side of 
(\ref{eq:Vjn-integral}) and (\ref{eq:qmn-integral}).  Thus  
$V_{jn}$ and $q_{mn}$, which are now functions of $a$ and $T$, 
can be expressed as second derivatives of the prepotential: 
\beqn  
    V_{jn} = \frac{1}{2 \pi i}\frac{\rd^2 \calF}{\rd a_j \rd T_n}, 
    \quad 
    q_{mn} = \frac{1}{2 \pi i}\frac{\rd^2 \calF}{\rd T_m \rd T_n}. 
\eeqn
Similarly, the matrix elements of the period matrix can be written 
\beqnarray
    \calT_{jk} = \frac{\rd^2 \calF}{\rd a_j \rd a_k}. 
\eeqnarray
Since $V_n$ should be the gradient vector of the potential 
$\calP_n$ in the $a$-space, we conclude that the potential 
$\calP_n$ can be written 
\beqn
    \calP_n = \frac{1}{2 \pi i} \frac{\rd \calF}{\rd T_n}.   
\eeqn
Thus all relevant quantities turn out to be written as 
derivatives of the redefined prepotential.  It is remarkable 
that the last equation resembles the hypothetical Hamilton-Jacobi 
equation of Losev et al. \cite{bib:Losev-etal}  In our case, 
however, $\calP_n$ is a function of both the Coulomb moduli 
AND the Whitham time variables, the latter enter from $dS$ 
through the integral formula, and we do not know how to convert 
it into a function of $a_j$ and $a^D_j = \rd \calF / \rd a_j$ 
like the Hamiltonians of Losev et al.

The redefined prepotential is a purely theoretical 
backbone, and not very suited for explicit calculations.  
Integral formulae, such as (\ref{eq:Vjn-integral}), 
(\ref{eq:qmn-integral}) and an integral representation of 
$\calP_n$ derived from (\ref{eq:DefEqF-Tn}), are more 
convenient.  These integral formulae remain valid even if 
the Whitham time variables are returned to the ``departure time'' 
$T = T_{\rm SW}$ where $dS$ is equal to $dS_{\rm SW}$; 
everything can be thereby calculated in terms of the original 
geometric data on the spectral curve.  Calculations of 
contact terms are thus eventually reduced to residue calculus.  
This is enough for understanding the four-dimensional problem.

Nevertheless, special families of Whitham deformations can 
possess some significant implications.  Gorsky et al. 
\cite{bib:Gorsky-etal-RGE} indeed presented such an example 
in the case of $G = SU(N)$ and $N_f = 0$.  
As they remarked, their Whitham deformations exhibit 
a remarkable similarity with two-dimensional topological 
Landau-Ginzburg models \cite{bib:DVV-LG}.  
The construction starts from the meromorphic differentials 
\beqn
    d\Omegahat_n = \Bigl( P(x)^{n/N} \Bigr)_+ d\log z, 
\eeqn
where $(\cdots)_+$ means the polynomial part of the 
Laurent expansion of $P(x)^{n/N}$ at $x = \infty$. 
$d\Omegahat_1$ is nothing but the differential 
$dS_{\rm SW}$. Now introduce a set of time variables 
$\Ttilde_n$ and consider the differential 
\beqn
    dS = \sum_{n=1}^\infty \Ttilde_n d\Omegahat_n. 
\eeqn
Because of a reason (see below), we have to distinguish 
between these time variables $\Ttilde_n$ and the previous 
ones $T_n$. The period integrals 
\beqn
    a_j = \oint_{\alpha_j} dS. 
\eeqn
define a function of the Coulomb moduli $u = (u_2,\cdots,u_N)$ 
and the time variables $\Ttilde_1,\Ttilde_2,\cdots$.  One can 
prove, by a standard method in Seiberg-Witten geometry, that 
the Jacobian matrix $\det(\rd a_j / \rd u_k)$ does not vanish 
in a neighborhood of, say, the ``Seiberg-Witten point'' 
$\Ttilde_n = \delta_{n,1}$.  Therefore the above relation 
can be solved for the Coulomb moduli as 
$u_j = u_j(a, \Ttilde_1,\Ttilde_2,\cdots)$.  This gives a 
deformation family of the spectral curve $\Sigma$.  Now 
modify the meromorphic differential $d\Omegahat_n$ into 
\beqn
    d\Omegatilde_n = d\Omegahat_n 
       - \sum_{j=1}^{N-1} \Bigl( \oint_{\alpha_j} 
         d\Omegahat_n \Bigr) d\omega_j, 
\eeqn
and write $dS$ in the form
\beqn
    dS = \sum_{n=1}^\infty \Ttilde_n d\Omegahat_n 
       + \sum_{j=1}^{N-1} a_j d\omega_j. 
\eeqn
One can then derive, by the method of Itoyama and Morozov 
\cite{bib:Itoyama-Morozov}, the Whitham equations 
\beqn
    \rd_{\Ttilde_n} dS|_{z={\rm const.}} = d\Omegatilde_n, \quad 
    \rd_{a_j} dS|_{z={\rm const.}} = d\omega_j. 
\eeqn

This Whitham deformation family is slightly distinct from 
those that we have considered.  A natural choice of local 
coordinates $z_\pm$ at $P_\infty^\pm$ is the following: 
\beqn
    z_+ = z^{-1/N}, \quad z_- = z^{1/N}. 
\eeqn
The polynomials $f_n^\pm(z_\pm^{-1})$ giving the singular part 
of $d\Omegatilde_n$ at $P_\infty^\pm$ can be written 
\beqnarray
    f_n(z_+^{-1}) &=& \frac{N}{n} \Lambda^n z_+^{-n} + \cdots, 
    \nonumber \\
    f_n(z_-^{-1}) &=& - \frac{N}{n} \Lambda^n z_-^{-n} + \cdots. 
\eeqnarray
The difference lies in the tail part ``$\cdots$''.  This part 
vanishes for $n < 2N$, but remains for $n \ge 2N$, and the 
coefficients of this part are NOT a numerical constant but 
a polynomial of the moduli $u_j$'s.  This is the reason that 
we changed the notation of the time variables.  This somewhat 
strange situation forces us a careful treatment of the 
prepotential.  For instance, when differentiating contour 
integrals like those in (\ref{eq:DefEqF-Tn}) against $T_m$ 
and $a_k$, we use the fact that the part of 
$f_n^\pm(z_\pm^{-1})$ may be considered constant; 
this is not permitted in the above example.

Extending this example to other gauge groups, such as 
$SO(2N)$, is an interesting problem.  We shall consider 
this issue elsewhere.

\section{Conclusion}

Inspired by the work of Mari\~no and Moore 
\cite{bib:Marino-Moore}, 
we have proposed an extension of the blowup formula 
with more than one 2-observables of the form $I_n(B)$ 
supported on the exceptional divisor $B$. Our strategy 
is simply to replace the single-time tau function of 
the Toda lattice hierarchy (in the sense of Mari\~no 
and Moore) by a multi-time tau function.  The time 
variables $t_n$ are interpreted as the coupling constants 
for the insertion of $I_n(B)$.  The tau function is 
an algebro-geometric tau function comprised of a Gaussian 
factor $e^{Q(t_1,t_2,\ldots)}$ and a theta function 
$\Theta_{\gamma,\delta}(\sum_n t_n V_n \mid \calT)$. 
The coefficients of the Gaussian part are identified 
to be the contact terms of the 2-observables.  We have 
partly confirmed the validity of our proposal by showing 
that the tau function possesses a desirable modular 
transformation property.  

This proposal has been further examined in the language 
of the Whitham equations that underlie the integrable 
hierarchy.  We have shown that the contact terms, as well 
as other relevant quantities, are written in the 
form of derivatives of the prepotential with respect to 
the Whitham time variables $T_n$.  This also clearly 
explains why the RG equation takes place in the description 
of contact terms of the quadratic Casimir $u_2$. 

These observations should be further checked in a 
field-theoretic language.  The superfield formalism 
of Losev et al. \cite{bib:Losev-etal} will provide 
a suitable framework for this purpose.  

It is remarkable that the integrable hierarchy and 
the Whitham hierarchy are both linked with two-dimensional 
field theories.  The integrable hierarchy (the Toda lattice 
hierarchy in the case of $G = SU(N)$ and $N_f = 0$) is 
related to massless free fermions on the spectral curve 
\cite{bib:tau-fermion}. 
(This is also an implicit message from the work of 
Gorsky et al. \cite{bib:Gorsky-etal-RGE}  They used 
the Szeg\"o kernels, which are correlation functions 
of free fermions.)  
The modular transformation property of the tau function 
is physically a consequence of conformal invariance of 
the two-dimensional massless free fermion theory. 
Meanwhile, as recent attempts at an analogue of the WDVV 
equations \cite{bib:SW-WDVV} suggest, the Whitham hierarchy 
are related to two-dimensional topological CFT's, 
in particular, topological Landau-Ginzburg models.  

These two-dimensional structures deserve to be studied 
in more detail. Of particular interest will be to examine, 
from our point of view, the mirror-like structure pointed 
out by Ito and Yang \cite{bib:SW-WDVV}.  

\section*{Acknowledgements}

The author is grateful to Toshio Nakatsu for discussions. 
He also thanks Masahiko Saito for invitation to his 
workshop held in Kinosaki, a beautiful hot spring town, 
where this work was started.  This work is partly 
supported by the Grant-in-Aid for Scientific Researches 
(No. 09304002), the Ministry of Education, Science and 
Culture, Japan.

\end{document}